# Nonlinear dynamic strand model with coupled tension and torsion

**S. Goyal[1] and N. C. Perkins**
University of Michigan, Mechanical Engineering
2350 Hayward, Ann Arbor, Michigan-48109-2125
Email. sgoyal@umich.edu

Marine cables under low tension and torsion form loops and tangles. In this context, loops are sometimes referred to as "hockles" and these can lead to damage of mechanical cables or signal attenuation in fiber optic cables [1]. The same type of deformation is observed in long, flexible biomolecules such as DNA which must "supercoil" in order to fit within the small confines of the cell nucleus [2]. Supercoiling is also known to play a critical role in gene expression [3, 4] as it can regulate the twist of the duplex. This analogy between cable "hockling" and DNA "supercoiling" is well-recognized in the literature [5-9].

Previous studies of cable hockling and DNA supercoiling have employed nonlinear *equilibrium* rod theories [9, 10] to compute the looped geometries. Special attention has also been given to understanding loop formation in generic elastic rods [11]. These studies all provide the means to compute three-dimensional equilibrium states and to assess stability of these states. It is well-known that some states become unstable under certain loading conditions; consider, for instance, the loop formation instabilities noted in [10, 11] and the opposite "pop-out" instabilities noted in [11, 12]. These instabilities lead to large dynamic responses which can also produce nonlinear transitions to more energetically favorable equilibria. Few studies, however, have investigated these dynamic responses. The dynamic rod model in [13] is used to study quasi-static response, while that in [14, 15] also captures the highly dynamic responses triggered by instabilities. The effects of dynamic self-contact are specifically captured in the model of [16, 17].

The purpose of the present study is to extend prior work on dynamic responses by including the important effects due to inertia and kinematic coupling of tension and torsion. We begin by reviewing a *dynamic* Kirchhoff rod model of a "strand" (length of cable or DNA) in the form of a $12^{th}$ order partial differential equation system. Numerical solutions reveal the effects of dynamics on clamped strands due to slow movements of their ends.

An example result is shown in Figure 1 below which corresponds to a twisted and compressed strand. The strand is initially twisted and then clamped at its ends. The twisted strand is then subject to compression (moving one end slowly towards the other) which ultimately produces the three-dimensional geometry shown here. We note that inertia plays a critical role in this process, particularly under loading conditions that pass through an instability. This fact is illustrated in Figure 2.

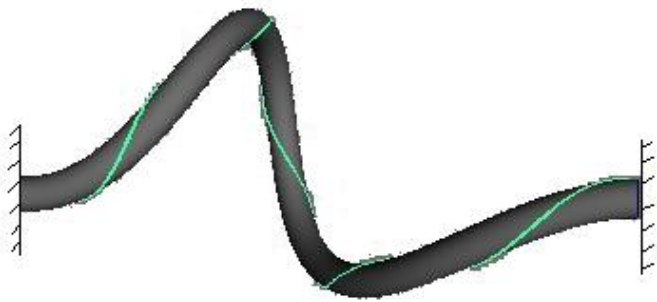

**Figure 1**: Example simulation of a twisted clamped strand.

Figure 2 summarizes a series of solutions for a twisted strand whose ends are slowly approaching. In particular, the resultant end load (along the tangent) is plotted as a function of end shortening where a shortening value of one corresponds to zero end separation. The solutions marked "statics" are those reported in [11] and are paths of equilibrium solutions initiated by distinct buckling "modes." The dynamic solutions capture an abrupt transition between these paths. This occurs when one path is energetically more "favorable" than the other. These predictions are also sensitive to the modeling of tension and torsion. For instance, if the tension is kinematically related to torsion, as expected in wire-rope constructions [18], then the dynamic transitions and the equilibrium paths are altered.

---
[1] Corresponding author

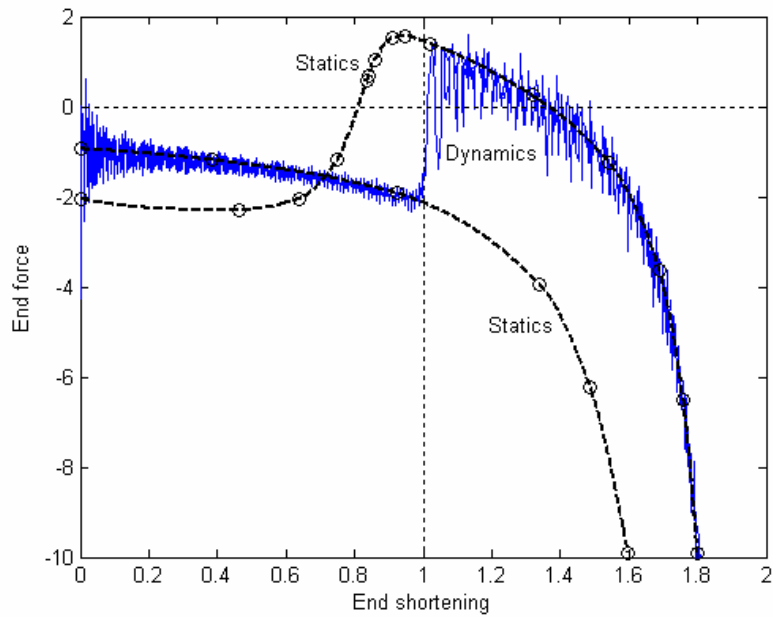

**Figure 2:** Computed end force as a function of end shortening for a twisted strand. End force is normalized with respect to fundamental Euler buckling load (negative in compression). The dynamic solution abruptly jumps between two possible sequences of equilibrium solutions [11].